\documentclass[  12pt]{article}
\usepackage{graphicx} 

\usepackage[margin=1.1in]{geometry}

\usepackage{enumerate}
\usepackage{natbib}
\usepackage{url} 
\usepackage{hyperref}

\usepackage{amsmath,amssymb,amsmath,amsfonts,amsthm}

\usepackage{tikz}
\usetikzlibrary{arrows,petri,shapes,automata,positioning}

\usepackage{caption}        
\usepackage{subcaption}     
\usepackage{multirow} 

\usepackage{algorithm}      
\usepackage{algpseudocode}  

\graphicspath{{Figures/}}

\title{Evaluating Randomness Assumption: A Novel Graph Theoretic Approach}
\author{Shriya Gehlot\\
    Operations and Decision Sciences\\ Indian Institute of Management Ahmedabad\\
        and \\
    Arnab Kumar Laha \\
    Operations and Decision Sciences\\ Indian Institute of Management Ahmedabad}
\date{26 June 2025}

\theoremstyle{plain}
\newtheorem{thm}{Theorem}[section]

\newtheorem{lemma}[thm]{Lemma}

\theoremstyle{definition}

\newtheorem{exmp}[thm]{Example}

\begin{document}

\def\spacingset#1{\renewcommand{\baselinestretch}%
{#1}\small\normalsize} \spacingset{1}

\maketitle

\begin{abstract}
Randomness or mutual independence is a fundamental assumption forming the basis of statistical inference across disciplines such as economics, finance, and management. Consequently, validating this assumption is essential for the reliable application of statistical methods. However, verifying randomness remains a challenge, as existing tests in the literature are often restricted to detecting specific types of data dependencies. In this paper, we propose a novel graph-theoretic approach to testing randomness using random interval graphs (RIGs). The key advantage of RIGs is that their properties are independent of the underlying distribution of the data, relying solely on the assumption of independence between observations. By using two key properties of RIGs—edge probability and vertex degree distribution—we develop two new randomness tests: the RIG-Edge Probability test and the RIG-Degree Distribution (RIG-DD) test. Through extensive simulations, we demonstrate that these tests can detect a broad range of dependencies, including complex phenomena such as conditional heteroskedasticity and chaotic behavior, beyond simple correlations. Furthermore, we show that the RIG-DD test outperforms most of the existing tests of randomness in the literature. We also provide real-world examples to illustrate the practical applicability of these tests.
\end{abstract}

\noindent%
\vfill

\newpage
\spacingset{1.4} 

\section{Introduction}
\label{intro}

Randomness plays a central role in empirical research and decision-making across a wide range of disciplines. Whether analyzing markets, evaluating policies, or modeling behavior, researchers often begin with the foundational assumption that data are drawn as independent realizations from a target population. This assumption that the observations represent a random sample of mutually independent and identically distributed variables is deeply embedded in statistical modeling literature. In economics, for example, methods assuming randomness are used to assess the variation of carbon price recommendations \citep{carbon_price} and effects of social interactions on group ties in an organization \citep{group_ties}. 
In finance, stock returns and asset price changes are usually assumed to be random \citep{time_series}.  In business management, randomness underlies practices such as quality control in manufacturing, randomized surveys in organizational behavior studies, and experimental designs in consumer research. 
Furthermore, widely used statistical tools such as regression analysis, goodness-of-fit tests, and one-sample t-test critically depend on the assumption of randomness in sampling.

Despite its widespread use, the randomness assumption is often not tested, which can lead to flawed conclusions when it is violated. 
Following \cite{miller}, consider a one-sample t-test applied to data $y_i$ presumed to be drawn from a $N(\mu,\sigma^2)$ distribution. Under the assumption that the data is random, a t-statistic value of 2.1 yields a p-value of 0.036, leading to the rejection of the null hypothesis at the 5\% level of significance. However, if the data actually exhibits serial correlation, such that $Cov(y_i,y_{i+1}) = \rho\sigma^2$ and $Cov(y_i,y_{i+j}) = 0, j \neq 0,1$, the t-statistic would asymptotically follow a $ N(0,1+2\rho)$ distribution instead of $N(0,1) $.  Even with a moderate correlation of $\rho=1/3$, the variance inflates to 5/3, significantly deviating from 1. Consequently, the same t-value of 2.1 now yields a p-value of 0.103, failing to reject the null hypothesis at the 5\% level of significance.  Thus, dependence in data can significantly affect the p-value, potentially resulting in erroneous conclusions. 

The implications of not testing randomness are evident in various domains. For instance, \cite{baris} uses the independent two-sample t-test to compare the views of leaders of Pacific Island Countries and the UN General Assembly on climate change. While doing this, the authors do not check for randomness in the data, and when we apply the randomness tests (Runs test \citep{Runs} or the tests developed in this paper) to it, the results show that the data is not random in nature. This discovery raises important concerns about the correctness of the conclusions made using the t-test in the paper.

\begin{figure}[t]
	\centering 
	\begin{subfigure}{0.45\textwidth}
		\centering\includegraphics[scale=0.4]{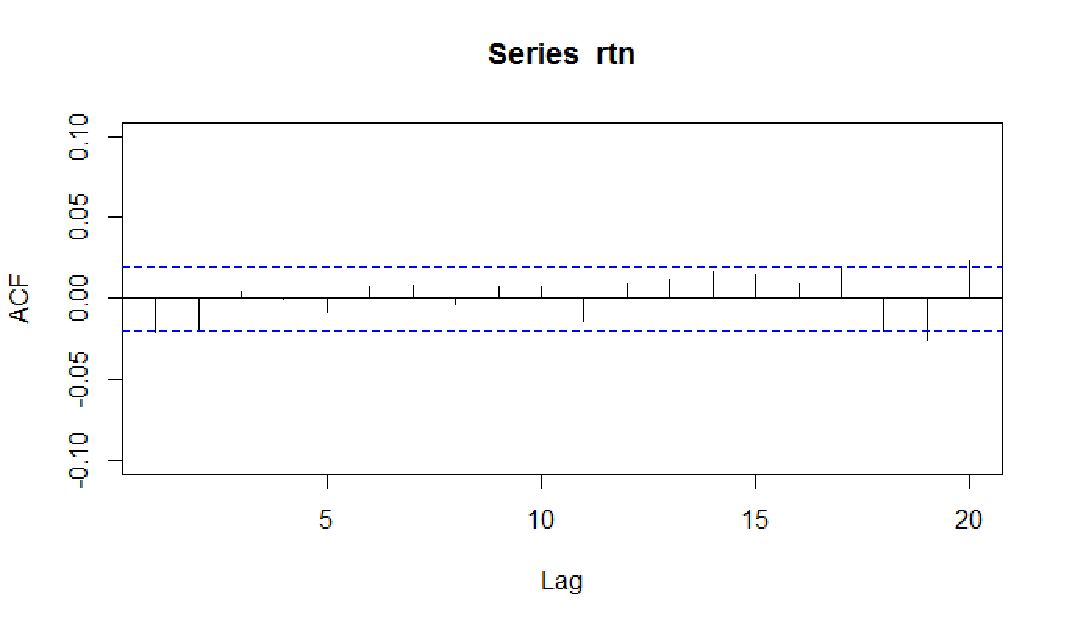}
		\caption{ACF of simple returns}
            \label{fig:IBM_a}
	\end{subfigure}\hfil 
	\begin{subfigure}{0.45\textwidth}
		\centering\includegraphics[scale=0.4]{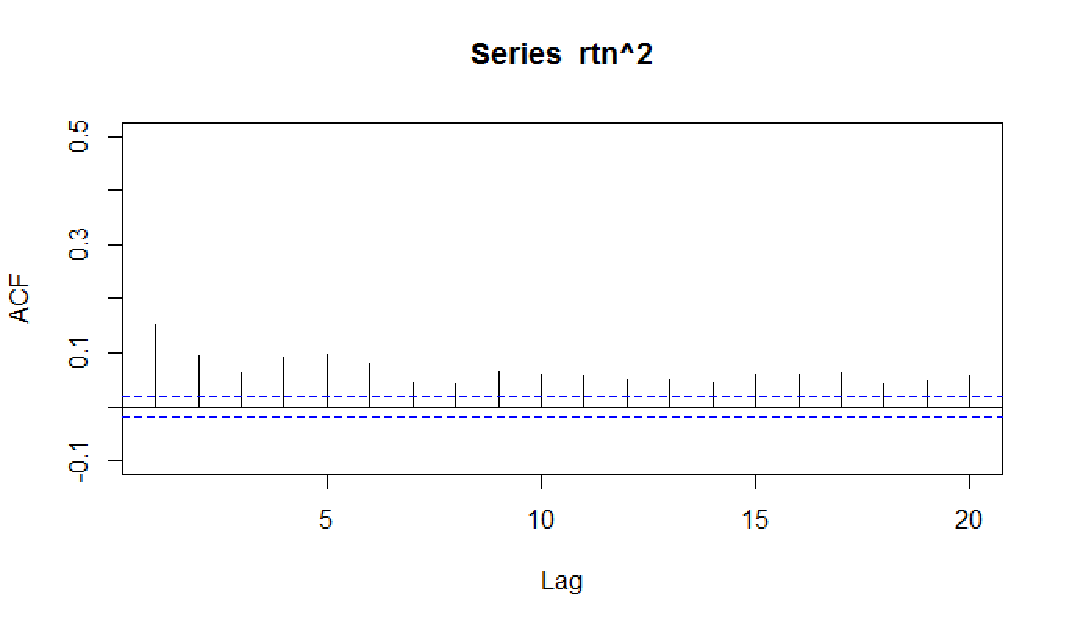}
		\caption{ACF of squared returns}
            \label{fig5:IBM_b}
	\end{subfigure}\hfil 
        \begin{subfigure}{0.45\textwidth}
		\centering\includegraphics[scale=0.4]{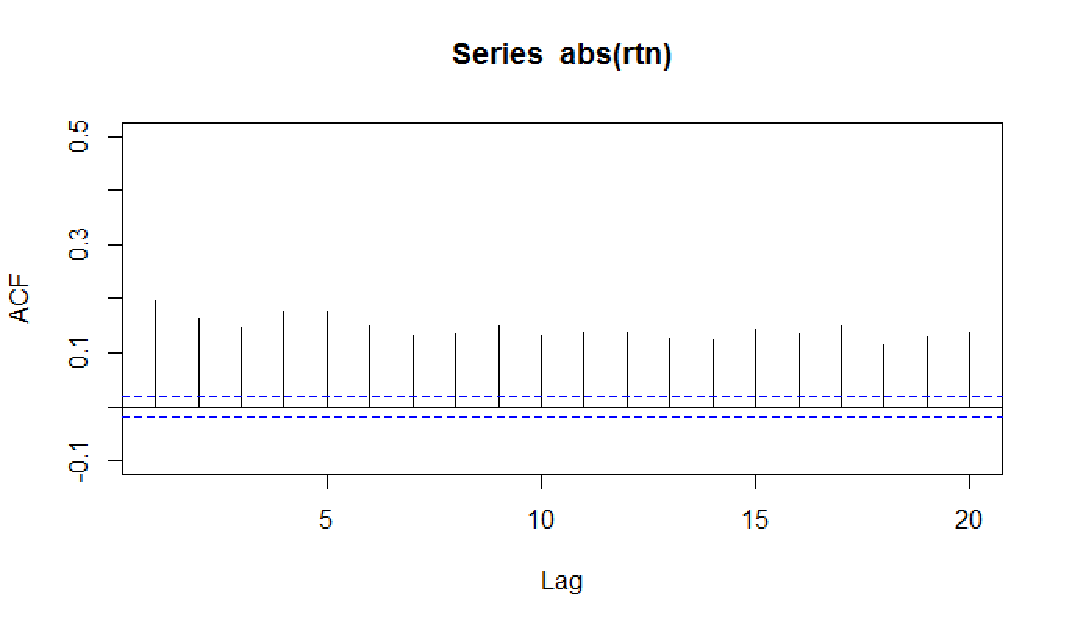}
		\caption{ACF of absolute returns}
            \label{fig:IBM_c}
	\end{subfigure}\hfil 
         \begin{subfigure}{0.45\textwidth}
		\centering\includegraphics[scale=0.4]{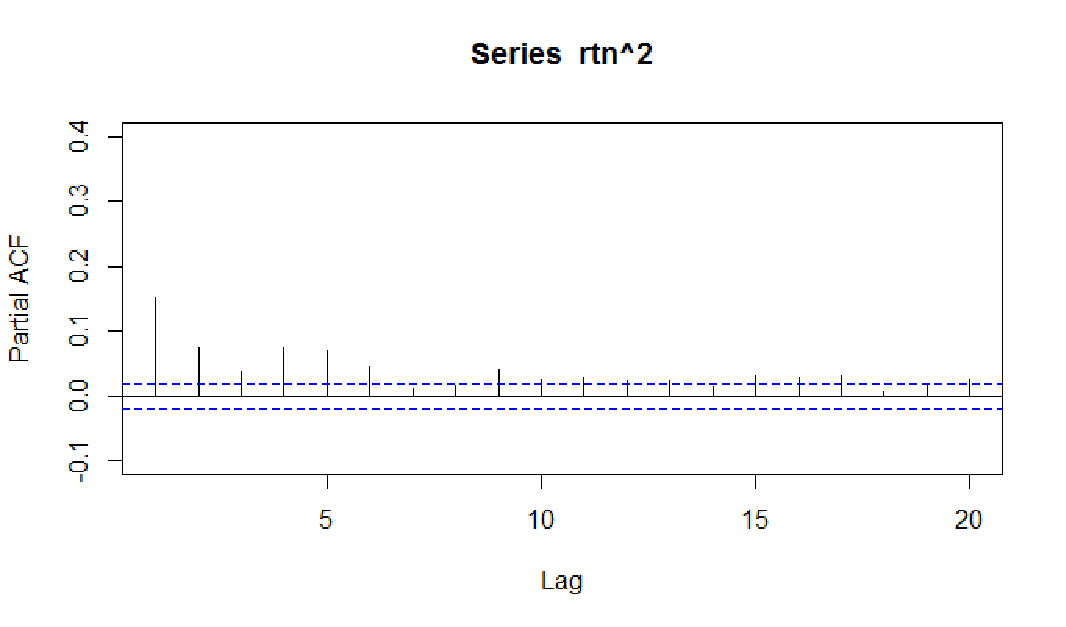}
		\caption{PACF of squared returns}
            \label{fig:IBM_d}
	\end{subfigure}\hfil 
	\caption{Sample ACF and PACF of various functions of daily simple returns of IBM stock.}
	\label{fig:IBM}
\end{figure}

Consider another example from the field of Finance - the daily simple returns of IBM stock. The autocorrelation function (ACF) plot of these returns (Fig. \ref{fig:IBM_a}) suggests no significant autocorrelation in the data. When the Runs test is applied to the data, it also shows no evidence against randomness at the 5\% level of significance. However, if we look at the ACF plots of the squared and absolute returns (Figs. \ref{fig5:IBM_b} and \ref{fig:IBM_c}), they show the possibility of the presence of dependence in the data. This dependence may arise from volatility, which fluctuates over time in ways that are not immediately observable. Notably, the Runs test fails to detect this type of dependence. This example underscores the limitations of current randomness tests in capturing more complex forms of data dependencies. We will revisit this example in Section \ref{real_data}.

These examples highlight two key issues in the existing literature. First, there is a lack of attention to the possibility of non-randomness in many studies, which jeopardizes the validity of their conclusions. Second, the scope of current randomness tests is often limited, as they tend to address only specific forms of dependence, leaving more complex dependencies undetected. In this paper, we aim to address these issues.

Although the randomness of observations is a fundamental assumption in many inferential problems, the tools available to test this assumption remain surprisingly limited. One of the most commonly used tests, aside from the Runs test, is the von Neumann Ratio test \citep{vonNeumann}. However, this test requires the data to follow a Gaussian distribution, and therefore cannot be applied to wide range of scenarios where the data are non-Gaussian. Other tests, such as the Ljung-Box test \citep{Ljung_Box}, the Breusch–Godfrey test \citep{Breusch, Godfrey}, and the Durbin-Watson test \citep{D_W1, D_W2} detect only autocorrelation in the data, with the Durbin-Watson test being restricted to detecting lag-1 autocorrelation. In non-parametric contexts, the Runs test is the most commonly used method. However, its effectiveness decreases with smaller datasets, which are common in real-world applications where data collection is limited or costly. Furthermore, the Runs test detects only serial correlation, missing more complex forms of dependence such as conditional heteroskedasticity. For non-linear dependencies, the BDS test \citep{bds} is often used; however, \cite{kanzler} has shown that its performance is highly sensitive to parameter choices. Even with careful parameter selection, the BDS test often fails to maintain its level, even for larger sample sizes ranging from several hundred to a few thousand observations \citep{broock}. This raises significant concerns about the reliability of results derived from this test, further emphasizing the need for more randomness tests.

In this paper, we propose a new approach to developing non-parametric tests for checking randomness in data using \textit{random interval graphs (RIGs)} defined by \cite{3}. \cite{3} shows that the properties of the RIGs are independent of the choice of distribution of observations and solely depend on whether the observations are independent or not. We use this characteristic of RIGs to construct new tests for randomness. 
We show that these tests are reliable across a wide range of scenarios, including light-tailed and heavy-tailed distributions, dependencies extending beyond lag 1, as well as those involving conditional heteroskedasticity and chaotic processes. Our findings suggest that the proposed tests provide a versatile and effective tool for identifying randomness in diverse setups, thereby extending the applicability of randomness testing to a wider range of practical situations than previously possible.

This paper is organized as follows: Section \ref{RIG} provides an overview of RIGs, including key properties that serve as the foundation for developing the proposed randomness tests. In Section \ref{ToR}, we present two new tests to check for randomness. In Section \ref{RIG-EP}, we use the edge probability to introduce the \textit{RIG-Edge Probability} test, while in Section \ref{RIG-DD}, we use the vertex degree distribution of RIG to introduce the \textit{RIG-Degree Distribution} test for assessing randomness. Section \ref{exp} presents extensive simulations to assess the performance of these tests, including their application to ARMA, ARCH-GARCH, and chaotic processes. Additionally, we compare our results with those obtained from the Runs and BDS tests. In Section \ref{real_data}, we apply our tests to two real-world datasets: the quarterly growth rate of US GNP and the daily simple returns of IBM. Section \ref{conclusions} summarizes the key findings and conclusions. 


\section{Random Interval Graphs}
\label{RIG}

In this section, we introduce the concept of \textit{random interval graphs} (RIGs) as described in \cite{3}, and explore their properties, which will serve as the foundation for proposing our new randomness tests.

To begin, we first define what an \textit{interval graph} is. 
A graph $G=(V,E)$ is called an interval graph if the vertices of $G$ correspond to the intervals on the real line and there is an edge between any two vertices if the corresponding intervals have a non-empty intersection \citep{2,3}. 

In a random graph, the presence of edges between vertices is determined by a random process. This randomness can model various real-world phenomena, making random graphs a versatile tool for studying complex systems. Following \cite{3}, let us consider a collection of $2n$ independent identically distributed (i.i.d.) continuous random variables, denoted as $X_1, X_2, \ldots, X_n, Y_1, Y_2, \ldots, Y_n$, where $X_i, Y_i $ follow distribution $F$ (cdf).
Then a random interval graph (RIG) is defined as a graph $G = (V, E)$ where the $i^{th}$ vertex $v_i$ in $G$ corresponds to the interval $I_i = [X_i,Y_i]$ if $X_i < Y_i$ or $I_i = [Y_i,X_i]$ if $Y_i < X_i$. Here we say that there is an edge between two vertices $v_i, v_j \in V$ if the corresponding intervals intersect i.e., if $I_i \cap I_j \ne \emptyset$. 

It is observed in \cite{3} that it is not necessary to rely on the assumption that $X_i$ and $Y_i$ follow a specific distribution $F$. Instead, it suffices to assume that the intervals have their endpoints as the numbers $1, 2, \ldots, 2n$, in some random order, where all the $(2n)!$ potential permutations are equally probable \citep{4}.
Since the properties of an RIG do not depend on the specified distribution $F$, without loss of generality, it can be assumed that $X_i,Y_i \sim$ Uniform (0,1) distribution. Thus, the properties of an RIG depend solely on the randomness (mutual independence) of the observations. This key observation forms the foundation for developing tests for randomness.

The properties of RIGs have already been studied in literature by \cite{3}. We will restrict our focus to two essential properties of the RIG that are pertinent to our test for randomness. The first property describes the likelihood of an edge connecting any two vertices within the RIG, while the second property provides insight into the distribution of a vertex's degree in the RIG.

Let $\mathcal{G}_n$ denote the set of all RIGs with $n$ vertices. Let $X_{ij} = 1$ if there is an edge between the $i^{th}$ and $j^{th}$ vertices of an RIG and $X_{ij}=0$ otherwise, $i \neq j$.

\begin{thm}[Scheinerman 1988]
\label{edge_prob}
    For any $G \in \mathcal{G}_n$, $P(X_{ij} = 1) = \frac{2}{3}$ for all $1 \leq i,j \leq n$.
\end{thm}

\begin{proof}
    Let $I_i = [a,b]$ and $I_j = [c,d]$ be intervals corresponding to vertices $v_i, v_j$ of an RIG, respectively. Then $I_i \cap I_j = \emptyset$, if (i) $a < b < c< d$, (ii) $a < b < d< c$, (iii) $b< a < c< d$,  (iv) $b< a < d< c$, (v) $c < d < a<b$, (vi) $c < d < b< a$, (vii) $d < c < a<b$, (viii) $d < c < b < a$. Therefore, $P(X_{ij} = 0) = \frac{8}{4!} = \frac{1}{3}$, as required. 
\end{proof}

Thus, Theorem \ref{edge_prob} tells us that $X_{ij} \sim Bernoulli (2/3)$. \cite{3} also establishes that the variables $X_{ij}$ are identically distributed but not independent. This result will be referenced frequently in subsequent sections. For clarity and ease of reference, we present a formal mathematical statement of this result below.

\begin{lemma}
    \label{lemma_edge_dept}
    Let $X_{ij}$ be as defined above. Then $X_{ij}$ is independent of $X_{kl}$ if and only if $\{i,j\} \cap \{k,l\} = \emptyset$.
\end{lemma}

The next result gives us the variation in the degree of a vertex within the RIG, the detailed proof of which is given in \cite{3}.

\begin{thm}[Scheinerman 1988]
    \label{deg_thm_linear}
    Let $G$ be an RIG with $n$ vertices, and $v \in V(G)$ has the degree $d(v)$. Then for fixed $x \in [0,1]$, if $x \geq \frac{1}{2}$
    $$\lim_{n \rightarrow \infty} P(d(v) \leq xn) = 1- (1-x)\frac{\pi}{2}$$
    and if $x < \frac{1}{2}$
    $$\lim_{n \rightarrow \infty} P(d(v) \leq xn) = 1- (1-x)\left\{ \frac{\pi}{2} - 2 \cos^{-1} \left[ \frac{1}{\sqrt{2-2x}} \right] \right\} - \sqrt{1-2x}.$$
\end{thm}

Building on Theorem \ref{deg_thm_linear}, we derive the following result for the joint distribution of two vertices in an RIG in Theorem \ref{2_arc_fix}. 
 
\begin{thm}
    \label{2_arc_fix}
    Let $d_1$ and $d_2$ be the degrees of any two vertices of $G \in \mathcal{G}_n$. Then,
    $$\lim_{n \rightarrow \infty}P(d_1 \leq xn,d_2 \leq xn) =  \left[ \lim_{n \rightarrow \infty}P(d_j \leq xn) \right]^2 =  \lim_{n \rightarrow \infty}P(d_1 \leq xn) \lim_{n \rightarrow \infty}P(d_2 \leq xn) ,$$
for $j=1,2$. 
\end{thm}

\begin{proof}
    Let $I_1 = [x_1,y_1]$ and $I_2 = [x_2,y_2]$ be two fixed intervals of $G \in \mathcal{G}_n$. Let $X_i^{(1)} = 1$ if $I_i \cap I_1 \neq \emptyset$ for $ 3 \leq i \leq n$, and $X_i^{(1)} = 0$ otherwise and $X_i^{(2)} = 1$ if $I_i \cap I_2 \neq \emptyset$ for $ 3 \leq i \leq n$, and $X_i^{(2)} = 0$ otherwise. Then $X_i^{(1)}, X_j^{(1)}$ are independent for $i \neq j$ and $X_i^{(2)}, X_j^{(2)}$ are independent for $i \neq j$. Let $X_{12} =1$ if $I_1 \cap I_2 \neq \emptyset$  and $X_{12} =0$ otherwise.  
Let $d_1 = \sum_{i \geq 3} X_i^{(1)} + X_{12}$ and $d_2 = \sum_{i \geq 3} X_i^{(2)} + X_{12}$ be the degree of the vertex corresponding to intervals $I_1$ and $I_2$ respectively. Define $r_j = \rho(I_j)$ to be the radius of the interval $I_j$ defined as $\rho(I_j) = \sqrt{a^2 + (1-b)^2}$ for $a = \min\{y_{2j-1},y_{2j}\}$ and $b = \max\{y_{2j-1},y_{2j}\}$.  Then $X_i^{(j)} \sim Bernoulli(p_j)$ for $3 \leq i \leq n$, $j=1,2$ where $p_j = P(X_i^{(j)}=1|x_j,y_j) =1-r_j^2$ \citep{3}.
Therefore $d_j \sim Bin(n-2,p_j)$ if $X_{12}=0$ and $d_j \sim Bin(n-2,p_j)+1$ if $X_{12}=1$. 

If $X_{12}=0$, then we can write $P(d_1 \leq xn,d_2 \leq xn|x_1,y_1,x_2,y_2) = P(Z_1\leq w_1, Z_2 \leq w_2)$ where $Z_j = \frac{d_j - (n-2)p_j}{\sqrt{(n-2)p_j(1-p_j)}}$ and 
$$ w_j =  \sqrt{n-2} \frac{x-p_j}{\sqrt{p_j(1-p_j)}} + \frac{2}{\sqrt{n-2}} \frac{x-p_j}{\sqrt{p_j(1-p_j)}} + \frac{1}{\sqrt{n-2}} \frac{2p_j}{\sqrt{p_j(1-p_j)}}.$$ 
Else, if $X_{12}=1$, then we can write $P(d_1 \leq xn,d_2 \leq xn|x_1,y_1,x_2,y_2) = P(Z_1\leq w_1, Z_2 \leq w_2)$ where $Z_j = \frac{d_j - (n-2)p_j-1}{\sqrt{(n-2)p_j(1-p_j)}}$ and 
$$ w_j =  \sqrt{n-2} \frac{x-p_j}{\sqrt{p_j(1-p_j)}} + \frac{2}{\sqrt{n-2}} \frac{x-p_j}{\sqrt{p_j(1-p_j)}} + \frac{1}{\sqrt{n-2}} \frac{2p_j-1}{\sqrt{p_j(1-p_j)}}.$$ 
This means, as $n \rightarrow \infty$, $w_j \rightarrow \infty$ if $x-p_j > 0$ and $w \rightarrow -\infty$ if $x-p_j < 0$ for $j=1,2$ irrespective of whether $X_{12}=0$ or $X_{12}=1$. Then, by the Central Limit Theorem, as $n \rightarrow \infty$,  $P(d_1 \leq xn,d_2 \leq xn | x_1,y_1,x_2,y_2) \rightarrow 1$ if $x-p_1 > 0$ and $x-p_2 > 0$ and $P(d_1 \leq xn,d_2 \leq xn | x_1,y_1) \rightarrow 0$ if $x-p_1 < 0$ or $x-p_2 < 0$. Therefore, 
$\lim_{n \rightarrow \infty} P(d_1 \leq xn,d_2 \leq xn | x_1,y_1,x_2,y_2) = I_{(x-p_1 > 0) \: \cap \: (x-p_2>0)}.$ 
Now, 
$$P(d_1 \leq xn,d_2 \leq xn)  = E(I_{d_1\leq xn \: \cap \: d_2 \leq xn}) = E_{x_1,y_1,x_2,y_2}[E(I_{d_1\leq xn  \: \cap \: d_2 \leq xn}|x_1,y_1,x_2,y_2)]$$
    Therefore, by the Bounded Convergence Theorem,
    \begin{align*}
        \lim_{n \rightarrow \infty} P(d_1 \leq xn,d_2 \leq xn)  & = E_{x_1,y_1,x_2,y_2} \left[ \lim_{n \rightarrow \infty}E(I_{d_1\leq xn \: \cap \: d_2 \leq xn}|x_1,y_1,x_2,y_2) \right] \\
        & = E_{x_1,y_1x_2,y_2} (I_{x-p_1>0 \: \cap \: x-p_2 >0}) = P(x-p_1 > 0 \: \cap \: x-p_2 >0)\\
        & = P(r_1^2 > 1-x \: \cap r_2^2 > 1-x)
    \end{align*}

By Proposition 3.3 in \cite{3}, $P(r_j^2 >1-x), j=1,2$ depends only on $x$.
Therefore, both of these events do not depend on each other. Thus $\lim_{n \rightarrow \infty} P(d_1 \leq xn,d_2 \leq xn )$ $ = P(r_1^2 > 1-x) P(r_2^2 > 1-x) =P(d_1 \leq xn) P(d_2\leq xn)$.
\end{proof}

These results will be used in Section \ref{ToR} to build the randomness tests.


\section{Test for Randomness}
\label{ToR}


In Section \ref{RIG}, we saw that the characteristics like the probability of an edge between any two vertices and the vertex degree distribution of an RIG are invariant with respect to the choice of the distribution of observations. Thus, these properties depend solely on the fact that the observations are randomly generated (i.e., mutually independent). With this understanding, we devise two tests for checking the randomness of a given set of univariate observations. The first test is based on the fact that the probability of an edge between any two vertices is 2/3, while the second one is based on the vertex degree distribution of an RIG.



\subsection{RIG-Edge Probability Test}
\label{RIG-EP}


Suppose we are given a set of $m$ observations on the real line, represented as $y_1, y_2, \ldots, y_{m-1}, y_{m}$. Our objective is to build an RIG for this set of observations so that the properties of the RIG can be utilized in our analysis.  First, let $m=4n$. Then, from these observations, we create $2n$ intervals $I_j$, as $I_j$ = $[y_{2j-1}, y_{2j}] $ if $y_{2j-1} < y_{2j}$ or $I_j$ = $[y_{2j}, y_{2j-1}] $ if $y_{2j} < y_{2j-1}$ for $ 1\leq j \leq 2n$. These intervals correspond to $2n$ vertices of the RIG.
Next, we randomly form $n$ pairs of vertices from these $2n$ vertices, ensuring that each vertex is part of only one pair. We associate an edge between the two vertices in a pair if the intervals associated with them have a nonempty intersection (i.e. if $I_a$ and $I_b$ are the two intervals associated with the two vertices in a pair, then there is an edge between these two vertices if and only if $I_a \cap I_b \ne \emptyset$).   
Define $X_i = 1$ if the $i^{th}$ pair of vertices does not have an edge between them and $X_i = 0$ otherwise. 

Theorem \ref{edge_prob} tells us that if the observations are generated randomly, the probability that there exists an edge between any two vertices of an RIG is $\frac{2}{3}$. Thus, $X_i \sim Bernoulli (\frac{1}{3})$ for all $i$. Since the pairs are randomly formed and each vertex is part of only one pair, by Lemma \ref{lemma_edge_dept}, the random variables $X_i, 1 \le i \le n$ are mutually independent. Let $X = \sum_{i=1}^{n} X_i$ be the number of pairs of vertices whose associated intervals do not intersect. Then $ X \sim Bin (n,\frac{1}{3})$. We now use this fact to construct a test of randomness. 

Consider an arbitrary graph $G$ that is formed by following the above procedure. Let $p$ be the probability that an edge between two randomly chosen vertices in $G$ does not exist. If the observations are mutually independent, then  $p=\frac{1}{3}$ as noted above. Thus, to test for randomness, we test  
$H_0 : p=\frac{1}{3}$  against the alternative  $H_1: p \neq \frac{1}{3}.$
Let $\hat{p} = \frac{1}{n} \sum_{i=1}^{n} x_i$ be the proportion of pairs whose vertices are not joined by an edge.  Using $\hat{p}$ as the test statistic, we propose the test: Reject $H_0$ if $\left| \hat{p} - \frac{1}{3} \right| > c$, where $c$ is chosen such that the level of the test is $\alpha$.
It is well known that for large values of $n$, $\hat{p}$ follows a Normal$\left(\frac{1}{3}, \frac{2}{9n}\right)$ distribution using which the value of $c$ can be computed. 
For small values of $n$, an exact test (possibly randomized) can be constructed (refer to Chapter 9 in \cite{Book_Rohatgi}, pp. 429-432).  
Define the randomized test $\varphi(x_1, \ldots, x_n)$ as $\varphi(x_1, \ldots, x_n) =1$ if $\sum_{i=1}^{n} x_i \in K_1$,  $\varphi(x_1, \ldots, x_n) = \gamma$ if $\sum_{i=1}^{n} x_i \in K_2$ and  $\varphi(x_1, \ldots, x_n) =0$ otherwise, where $K_1, K_2 $ and $\gamma$ are chosen such that $E_{H_0}[\varphi] = \alpha$.
Therefore we reject $H_0$ with probability 1 if $\sum_{i=1}^{n} x_i$ takes values in $K_1 \subset \{0,\dots,n\}$ and with probability $\gamma \in (0,1)$ if $\sum_{i=1}^{n} x_i$ takes values in $K_2 \subset \{0,\dots,n\}$ and do not reject $H_0$ otherwise.

When the total observations $m$ is not a multiple of four, i.e., $m = 4n+k$ for $k =1,2,3$, we cannot use the above-mentioned method. To resolve this issue, one strategy involves omitting $k$ observations to achieve a dataset of length $4n$, enabling the use of the earlier method. If $k=3$, another ad-hoc approach is to select one observation randomly from the $m$ observations and repeat it in the dataset. 
Instead of the above ad-hoc approaches, we suggest a method that involves organizing the observations into groups of size $4n$ and repeating the testing process $k+1$ times to accommodate all the observations. For instance, if $k=3$, groups would be formed as follows: Group-1: $y_1, \dots, y_{4n}$, Group-2: $y_2, \dots, y_{4n+1}$, Group-3: $y_3, \dots, y_{4n+2}$ and Group-4: $y_4, \dots, y_{4n+3}$. We compute the test statistic $\hat{p}_i$ for the Group-$i$, $i=1, \ldots, k+1$. Thus, we have $k+1$ test statistics. Now, we apply the Benjamini-Hochberg method \citep{BY_correction} of multiple testing to obtain the adjusted p-values. If the least adjusted p-value is greater than the specified level, we do not reject $H_0$ in favour of $H_1$. We call this the \textit{RIG-Edge Probability  (RIG-EP) test}. Algorithm \ref{algo_RIG-EP} describes the RIG-EP test for randomness.

\begin{algorithm}
    \caption{RIG-Edge Probability Test}
    \begin{algorithmic}[1]
        \State  \textbf{Input:} A univariate series of $m$ observations $y_1,y_2,\dots, y_{m-1}, y_{m}$.
        \If {$m=4n$}
        \State Build an RIG with $2n$ vertices where $I_j = [y_{2j-1},y_{2j}]$ if $y_{2j-1} < y_{2j}$ or $I_j = [y_{2j},y_{2j-1}]$ if $y_{2j} < y_{2j-1}$, $1\leq j \leq n$.
	\State Choose $n$ pairs of vertices out of these $2n$ vertices without replacement, i.e., each vertex is there in only one pair. 
        \State Form an edge between the two vertices in a pair if the intervals associated with them have a nonempty intersection.
        \State Let $p$ be the probability that there does not exist an edge between any two vertices in an RIG.  For testing $H_0 : p=\frac{1}{3} \text{ vs } H_1: p \neq \frac{1}{3}$, compute $\hat{p}$, the proportion of pairs whose vertices are not joined by an edge. 
        \If {$n$ is large} \hspace{0.1cm}
         Reject $H_0$ if $|\hat{p} - \frac{1}{3}| > c$, where $c = \sqrt{\frac{2}{9n}}  z_{\frac{\alpha}{2}}$ for level $\alpha$.
        \Else \hspace{0.1cm}
        Reject $H_0$ with probability 1 if $n \hat{p} \in K_1$ and with probability $\gamma$ if $n \hat{p} \in K_2$, where $K_1,K_2,\gamma$ are determined based on the level $\alpha$
        \EndIf
        \ElsIf {$m = 4n+k, k =1,2,3$}
        \State Group observations into $k+1$ groups $G_1, \ldots, G_{k+1}$ of size $4n$ each, where $G_1=\{y_1, \ldots, y_{4n}\}$, $G_2= \{y_2, \ldots, y_{4n+1}\}$, \dots, $G_{k+1} = \{y_{k+1}, \ldots, y_{4n+k}\}$  
        \State Apply Steps 2-9 to each of the groups $G_i$ to obtain $k+1$ p-values.
        \State Apply the Benjamini and Hochberg procedure to obtain the adjusted p-values.
        \State  We do not Reject $H_0$ if the least adjusted p-value is $> \alpha$.
        \EndIf
    \end{algorithmic}
    \label{algo_RIG-EP}
\end{algorithm} 


\subsection{RIG-Degree Distribution Test}
\label{RIG-DD}


Another characteristic of the RIG that we will use to build a test for randomness is the vertex degree distribution. 
\cite{3} tells that the degree distribution of a vertex in an RIG remains the same regardless of the choice of distribution of observations. Consequently, upon fixing the vertex count, the theoretical degree distribution of a vertex of the RIG can be computed from Theorem \ref{deg_thm_linear}. Therefore, if we create the RIG for a given set of observations and the observations are mutually independent, the empirical and theoretical degree distributions should be aligned closely to each other. Thus, the similarity of the empirical and theoretical degree distributions could act as a measure to check for the randomness of the observations. 

\begin{figure}[t]
	\centering 
	\begin{subfigure}{0.3\textwidth}
		\centering\includegraphics[scale=0.35]{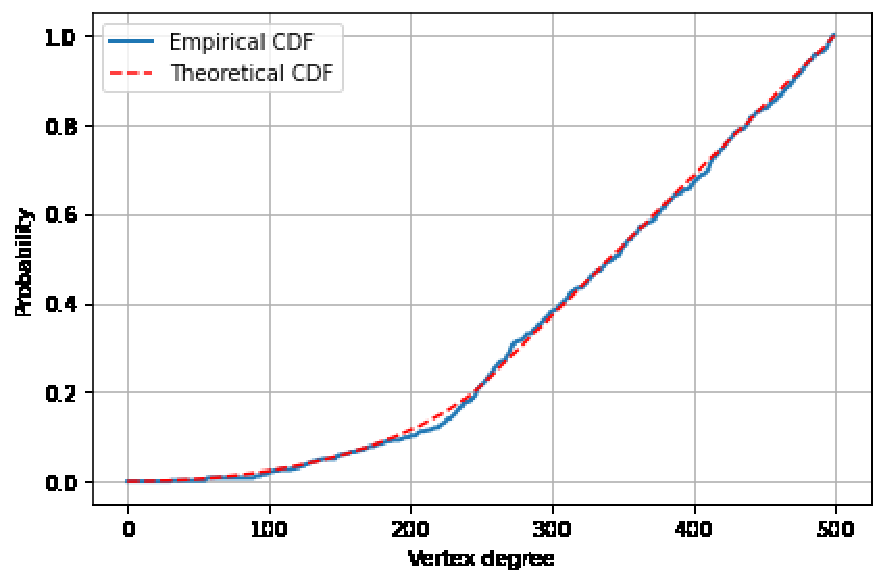}
		\caption{CDF Uniform[0,1]}
	\end{subfigure}\hfil 
	\begin{subfigure}{0.3\textwidth}
		\centering\includegraphics[scale=0.35]{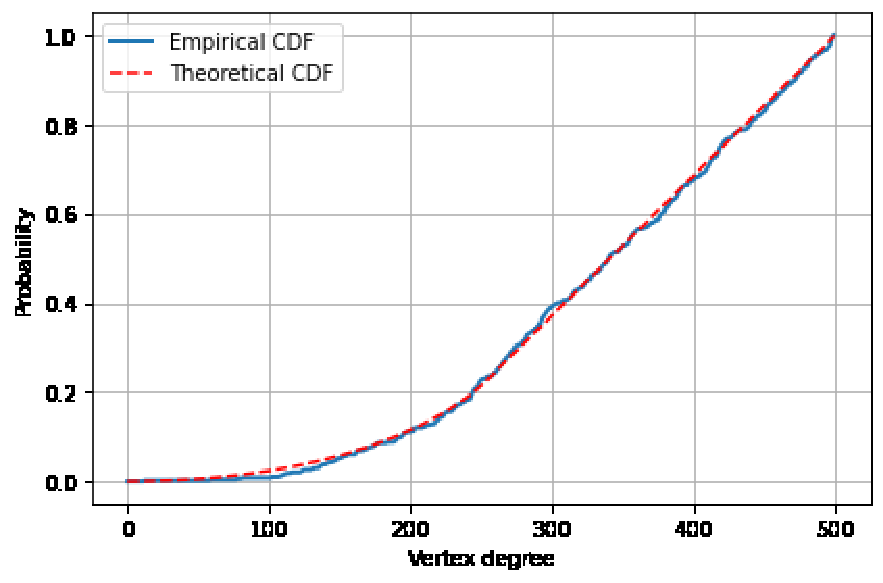}
		\caption{CDF Beta(2,3)}
	\end{subfigure}\hfil 
        \begin{subfigure}{0.3\textwidth}
		\centering\includegraphics[scale=0.35]{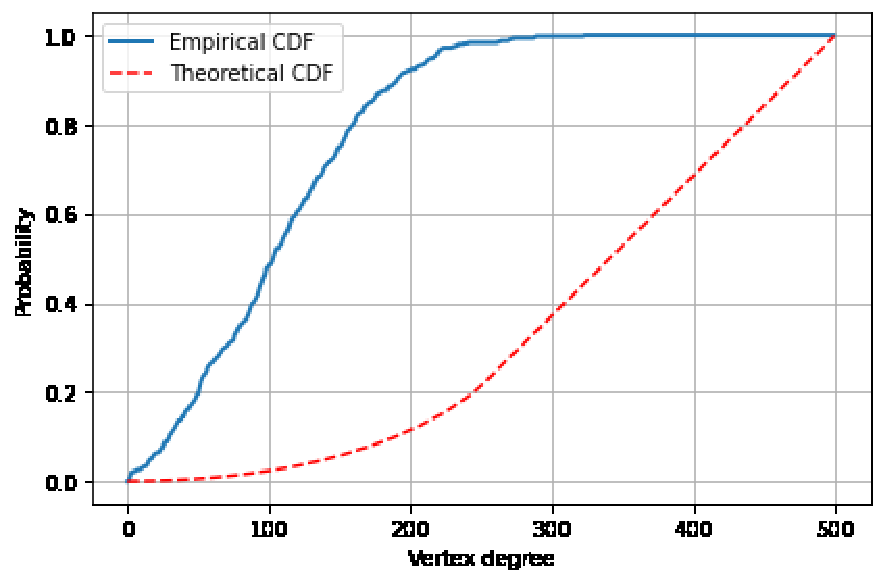}
		\caption{CDF AR(1) process}
	\end{subfigure}\hfil 
    \begin{subfigure}{0.3\textwidth}
		\centering\includegraphics[scale=0.35]{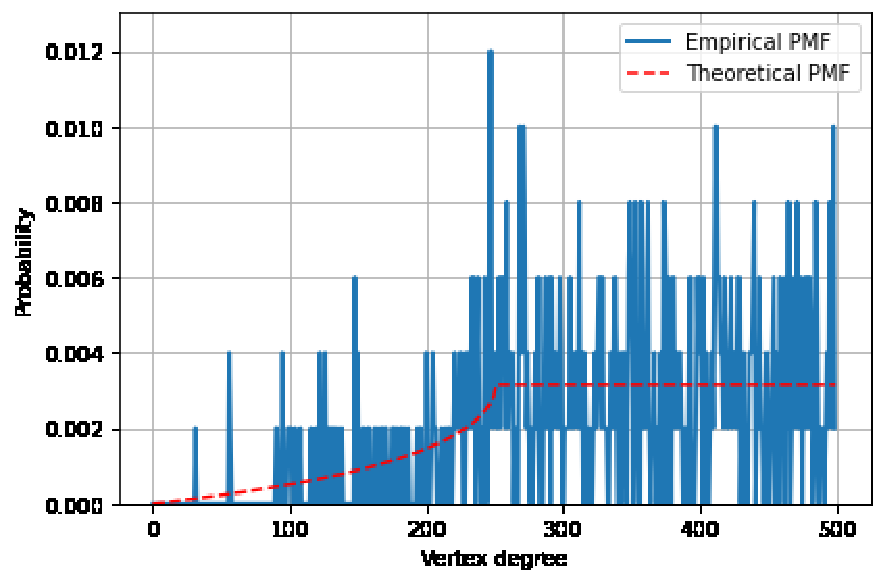}
		\caption{PMF Uniform[0,1]}
	\end{subfigure}\hfil 
	\begin{subfigure}{0.3\textwidth}
		\centering\includegraphics[scale=0.35]{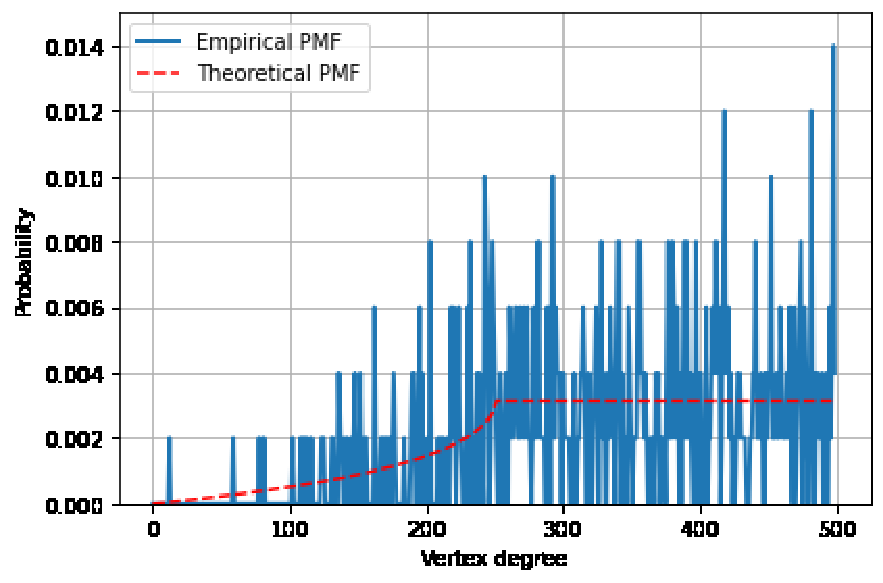}
		\caption{PMF Beta(2,3)}
	\end{subfigure}\hfil 
        \begin{subfigure}{0.3\textwidth}
		\centering\includegraphics[scale=0.35]{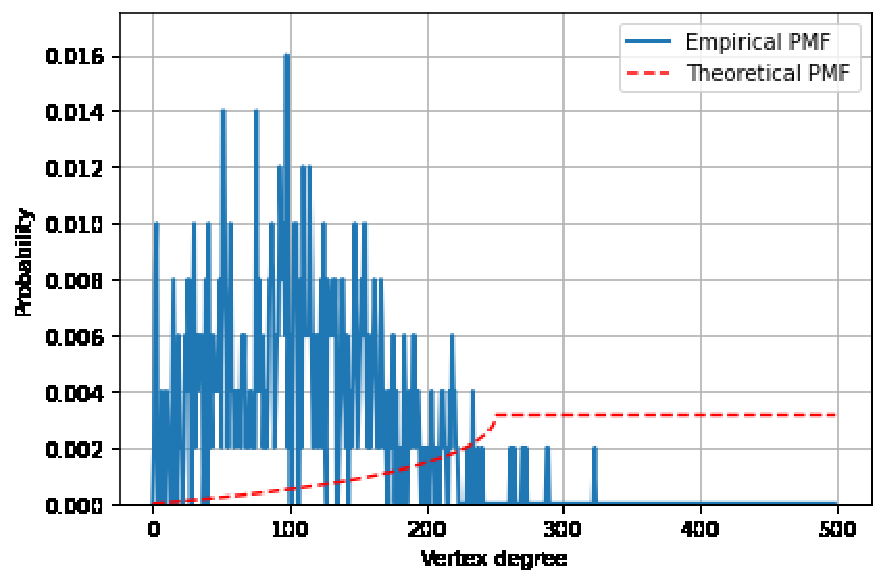}
		\caption{PMF AR(1) process}
	\end{subfigure}\hfil 
	\caption{Comparison between theoretical and empirical  vertex degree distributions of RIGs with 1000 observations when the data is generated randomly from Uniform[0,1] distribution in (a) and (d), randomly from Beta(2,3) distribution in (b) and (e), and AR(1) process in (c) and (f).}
	\label{fig_dd_exmpl}
\end{figure}

Figure \ref{fig_dd_exmpl} illustrates this idea with 1000 observations generated under three different scenarios: (i) randomly from a Uniform[0,1] distribution, (ii) randomly from a Beta(2,3), and (iii) from an Autoregressive (AR) process. The figure shows that when the observations are truly random, the empirical and theoretical degree distributions closely match. However, for the AR(1) process, where the observations exhibit dependence, a noticeable deviation of the empirical degree distribution from the theoretical degree distribution is observed.

Formally, suppose we are given a set of $m$ observations  $y_1,y_2,\dots,y_{m}$. First, let $m=2n$. Then we build an RIG $G$ corresponding to the given set of observations with $n$ vertices where the $j^{th}$ vertex corresponds to interval $I_j = [y_{2j-1},y_{2j}]$ if $y_{2j-1} < y_{2j}$ or $I_j = [y_{2j},y_{2j-1}]$ if $y_{2j} < y_{2j-1}$, $1\leq j \leq n$. Let $\hat{F_n}(x)  = \frac{1}{n}\sum_{j=1}^{n} I(d_j\leq xn)$ be the empirical degree distribution of this graph $G$ where $d_j$ is the degree of vertex $j$ and $F^*$ be the theoretical degree distribution of RIG with $n$ vertices.
Since the $d_j$'s are not i.i.d., we cannot claim that $\hat{F_n}$ converges to $F^*$ uniformly using the Glivenko-Cantelli theorem.

Note that, by Theorem \ref{deg_thm_linear}, we have $\lim_{n \rightarrow\infty} P(d_j \leq xn)= F^*(x)$, for every $1 \le j \le n$. Hence, for any given $\epsilon>0, \exists \: N_j$ such that $\forall \: n \geq N_j$, $|P(d_j \leq xn) - F^*(x)| < \epsilon$. Now, define $k(n)$ to be the number of $j$ for which $n \geq N_j$. 

\begin{thm}
    \label{Fn_convg}
    Let $\hat{F}_n(x)$ be as defined above and suppose $\lim_{n \rightarrow \infty} \frac{k(n)}{n} = 1$. Then $\hat{F}_n(x)$ converges in probability to $F^*(x)$ for every $x\in [0,1]$.
\end{thm}

\begin{proof}
    Let $S_1 = \{j \leq n : |P(d_j \leq xn) - F^*(x)|< \epsilon \}$ and $S_2 = \{j \leq n : |P(d_j \leq xn) - F^*(x)| \geq \epsilon \} $. Then $|S_1|= k(n)$ and $|S_2| = n-k(n)$. 

Let $t=xn$. Then as $n\rightarrow\infty, t \rightarrow \infty$. We have 
\begin{align*}
    E[\hat{F}_n(x)] &= \frac{1}{n}\sum_{i=1}^{n} E[ I(d_j\leq xn)]\\
    & = \frac{1}{n}\sum_{j=1}^{n} P(d_j \leq t)
\end{align*}
Now we can write the expression of $E[\hat{F}_n(x)]$ as
\begin{align*}
    E[\hat{F}_n(x)] &= \frac{1}{n}\sum_{j=1}^{n} \left[ P(d_j \leq t) -F^*(x) \right] + F^*(x)\\
    &= \frac{1}{n}\sum_{j\in S_1} \left[ P(d_j \leq t) -F^*(x) \right] +\frac{1}{n}\sum_{j\in S_2} \left[ P(d_j \leq t) -F^*(x) \right] + F^*(x)\\
    & \leq \frac{k(n)}{n} \epsilon + 2 \left( \frac{n-k(n)}{n} \right) + F^*(x).
\end{align*}
Since $\lim_{n \rightarrow \infty} \frac{k(n)}{n}=1$, we get $\lim_{n \rightarrow \infty}E[\hat{F}_n(x)] = F^*(x)$. Now, 
\begin{align*}
    Var &(\hat{F}_n(x)) = \frac{1}{n^2} Var \left( \sum_{j=1}^n I(d_j \leq t)\right) \\
    & = \frac{1}{n^2} \left( \sum_{j=1}^n Var \left(I(d_j \leq t)\right) + 2 \sum_{j=1}^n\sum_{i <j} Cov (I(d_i \leq t)I(d_j \leq t))\right) \\
    &= \frac{1}{n^2} \left( \sum_{j=1}^n P(d_j \leq t) (1-P(d_j \leq t)) + 2 \sum_{j=1}^n\sum_{i <j} \left[ P(d_i \leq t,d_j \leq t) -  P(d_i \leq t) P(d_j \leq t) \right] \right)\\
    &\leq \frac{1}{4n} + \frac{1}{n^2} \left( 2 \sum_{j=1}^n\sum_{i <j} \left[ P(d_i \leq t,d_j \leq t) -  P(d_i \leq t) P(d_j \leq t) \right] \right)
\end{align*}
Then by Theorem \ref{2_arc_fix}, we can conclude that $\lim_{n \rightarrow \infty} Var(\hat{F}_n(x)) =0$. Thus, by using Chebyshev's Inequality, we get the required result. 
\end{proof}

From this theorem, we can conclude that when the observations are mutually independent, $\hat{F}_n(x)$ converges to $F^*(x)$ in probability for every $x\in [0,1]$.  We now use this property to construct a test for randomness based on the degree distribution. 
We base the test on a Hellinger-like distance $D$ between the theoretical and empirical cdfs. We define $D = \frac{1}{\sqrt{2}} \sum_{i=1}^n \left( \sqrt{\hat{F}_n(\frac{i}{n})} - \sqrt{F^*(\frac{i}{n})}\right)^2$ as our test statistic. We define the rejection rule as: Reject the null hypothesis that the observations are random (mutually independent) if the value of $D$ is greater than some threshold value $C_\alpha$ chosen based on the level of significance $\alpha$. 
The null distribution of $D$ seems to be intractable to derive theoretically. Hence, we take recourse to simulation to determine the value of $C_\alpha$. 

Notice that here, we choose the distance $D$ instead of the Hellinger distance because the pmf of the vertex degree takes a constant value after $x \geq 1/2$. This makes the Hellinger distance less sensitive to capturing the deviation from the theoretical degree distribution as compared to distance $D$. 

When $m$ is not a multiple of two, one way to test for randomness is to randomly remove one observation, reducing the dataset to size 
$2n$ so that we can apply the method described above. However, since the result depends on the observation removed, we suggest an alternative procedure similar to the one used for the RIG-EP test. For $m=2n+1$, we create two groups: Group-1: $y_1, \dots, y_{2n}$  and Group-2: $y_2, \dots, y_{2n+1}$. We compute the test statistics $D_1$ and $D_2$ for these two groups separately.  We now reject the null hypothesis of randomness if any of the test statistics is greater than $C_{\alpha/2}$. 

We call this test the \textit{RIG-Degree Distribution (RIG-DD)} test. 
Algorithm \ref{algo_RIG_DD} describes the RIG-DD test for randomness.

In this paper, to calculate the value of $C_\alpha$, we generate 1000 random series of length $2n$ from the Uniform[0,1] distribution and compute test statistic values $D^{(1)}, D^{(2)}, \dots, D^{(1000)}$ for each of these generated random series. 
Then, for a test at a level of significance $\alpha$, $C_\alpha =100(1-\alpha)^{th}$ percentile of $D^{(i)}$. Algorithm \ref{algo_thrsld} describes the method of calculation of $C_{\alpha}$ through simulation. Table \ref{tab_cutoff_linear} in the Appendix gives the values of $C_\alpha$ for 10\%, 5\%, and 1\% levels of significance for various values of $m$. 

\begin{algorithm}
    \caption{RIG-Degree Distribution Test}
    \begin{algorithmic}[1]
        \State \textbf{Input:} A series of observations $y_1,y_2, \dots, y_{m}$.
        \If {$m=2n$}
        \State  Build an RIG with $n$ vertices where $I_j = [y_{2j-1},y_{2j}]$ if $y_{2j-1} < y_{2j}$ or $I_j = [y_{2j},y_{2j-1}]$ if $y_{2j} < y_{2j-1}$, $1\leq j \leq n$.
	\State Compute $\hat{F}_n$, the empirical degree distribution for the RIG obtained above. 
	\State Compute distance $D = \frac{1}{\sqrt{2}} \sum_{i=1}^n \left( \sqrt{\hat{F}_n(\frac{i}{n})} - \sqrt{F^*(\frac{i}{n})}\right)^2$, where
    $F^*$ is the theoretical degree distribution for an RIG with $n$ vertices.
        \State Reject null hypothesis of randomness if $D>C_\alpha$, where $C_\alpha$ determined based on level $\alpha$. An approximate value of $C_\alpha$ can be computed using Algorithm \ref{algo_thrsld}.
        \ElsIf {$m = 2n+1$}
        \State Group observations into 2 groups $G_1=\{y_1, \ldots, y_{2n}\}$, $G_2= \{y_2, \ldots, y_{2n+1}\}$ of size $2n$ each.  
        \State Apply Steps 2-6 separately to groups $G_1$ and $G_2$ to obtain two test statistics $D_1$ and $D_2$.
        \State Reject null hypothesis of randomness if either of $D_1$ or $D_2$ is greater than $C_{\alpha/2}$.
        \EndIf
    \end{algorithmic}
    \label{algo_RIG_DD}
\end{algorithm}

\begin{algorithm}
    \caption{Approximate Threshold Calculation}
    \begin{algorithmic}[1]
        \State Generate $k$ random series of length $2n$ from the Uniform[0,1] distribution (in this paper $k=1000$).
	\State Compute test statistic values $D^{(1)}, D^{(2)}, \dots, D^{(k)}$ for each of these generated random series.
       \State Compute the approximate threshold value of $C_\alpha$ to be the $ 100(1-\alpha)^{th}$ percentile of $D^{(i)}$'s for the level of significance $\alpha$.
    \end{algorithmic}
    \label{algo_thrsld}
\end{algorithm}

\section{Numerical Studies}
\label{exp}


In this section, we conduct experimental studies to examine the performance of our proposed RIG-EP and RIG-DD tests. We begin by providing examples that demonstrate the working of the RIG-EP and RIG-DD tests. Following this, we employ various simulation experiments for both large and small datasets to evaluate the efficacy of our tests in detecting randomness. 
While most existing tests primarily assess dependence in the form of serial correlation, we extend our analysis to consider situations involving more complex forms of dependence like that in ARCH and GARCH processes, where the observations are uncorrelated but dependent. Additionally, we evaluate the performance of our tests on chaotic processes, further demonstrating their versatility. We compare our results with two widely used tests, namely, the Runs test and the BDS test. Our findings demonstrate the broader applicability of our tests, showing that they provide a significant advantage over existing methods. 

For the RIG-EP test, we start by considering a truly random series to determine whether our test can accurately identify it. We generate this series of random numbers using RANDOM.ORG, which utilizes atmospheric noise as a source of randomness. Subsequently, we apply the RIG-EP test to this dataset, as outlined below.

\begin{exmp}
    Consider a set of 20 true randomly generated real integers between 1 and 100.
$$13, 73, 89, 78, 21, 51, 95, 8, 85, 80, 39, 43, 15, 33, 1, 11, 31, 87, 36, 19$$
We apply the RIG-EP test on it as given in Algorithm \ref{algo_RIG-EP}. This gives us an RIG with ten vertices as follows
$$[13,73], [78,89], [21,51], [8,95], [80,85],
[39,43], [15,33], [1,11], [31,87], [19,36].$$
Let us form five random pairs of intervals from these ten intervals as $$([78,89],[19,36]),([8,95],[80,85]),([13,73],[1,11]),([21,51],[39,43]),([15,33],[31,87]).$$
Here, the proportion of non-intersecting pairs of intervals ($\hat{p}$) is 0.4. Now as described in Section \ref{RIG-EP}, for $n=5$, the values of parameters are $K_1 = \emptyset, K_2 =\{0,5\}$, and $\gamma = 0.368$ for $\alpha = 0.05$. Therefore, $\varphi(x) = 0$, for this sample. Thus, we do not reject the null hypothesis that the observations are random at the 5\% level of significance.
\end{exmp}

To understand the workings of the RIG-DD test, we present the following example involving 1000 observations. 

\begin{exmp}
    Consider three distinct sets of 1000 observations generated as follows: (i) randomly from a Uniform[0,1] distribution, (ii) randomly from a Beta(2,3) distribution, and (iii) from an AR(1) process with $\rho=0.9$. We apply the RIG-DD test, as outlined in Algorithm \ref{algo_RIG_DD}, to each of these datasets. This involves calculating the theoretical degree distribution for an RIG with 500 vertices and comparing it to the empirical degree distributions obtained from these three cases.
    The distributions are illustrated in Figure \ref{fig_dd_exmpl}. When we generate the observations randomly from the Uniform and Beta distributions (cases (i) and (ii)), the empirical degree distributions closely resemble the theoretical degree distribution. However, in the case of non-random observations generated from the AR(1) process (case (iii)), the empirical degree distribution deviates significantly from the theoretical distribution. Specifically, the Hellinger distances between the theoretical degree distribution and the empirical distributions for cases (i), (ii), and (iii) are 0.213, 0.203, and 6.638, respectively. According to Algorithm \ref{algo_thrsld}, the threshold value at the 5\% level of significance for  $m=1000$ is 0.6989. Consequently, at the 5\% level of significance, we fail to reject the null hypothesis of randomness for the data generated from the Uniform and Beta distributions. In contrast, we reject the null hypothesis for the data generated from the AR(1) series, indicating non-randomness at the 5\% level of significance.
\end{exmp}

To evaluate the statistical power of both RIG-EP and RIG-DD tests, we conduct a series of simulation experiments as outlined below. Additionally, we compare our results with those obtained from the Runs and BDS tests. For all simulations, we set the parameters of the BDS test with an embedding dimension of 4 and a threshold distance equal to 0.5 times the standard deviation of the data. 

We begin by evaluating the performance of our tests in a simple scenario where serial correlation is present in the form of an AR(1) process. 
We generate a series of $m$ observations from an AR(1) process, defined as $Y_t = \rho Y_{t-1} + e_t$, $ -1 < \rho< 1, e_t \sim N(0,1)$, using various values $\rho$ for (i) $m=4000$ and (ii) $m=1000$. Subsequently, we apply the RIG-EP, RIG-DD, Runs, and BDS tests to assess randomness at the 5\% level of significance. To check the performance of our tests, we replicate this procedure 1000 times and record the frequency of rejection of the null hypothesis of randomness. The results are presented in Table \ref{tab_AR(1)_linear}. For $m=4000$, observe that even for $\rho = 0.2$, both RIG-EP and RIG-DD tests exhibit high power (approximately 1). We also observe that as the value of $\rho$ increases, the power of our tests also increases. 
This trend is consistent even for $m=1000$. 

Next, we compare the performance of the RIG-EP and RIG-DD tests against the Runs and BDS tests.  Our analysis demonstrates that the RIG-EP and RIG-DD tests consistently outperform the BDS test across all values of $\rho$ and $m$. As can be seen from Table \ref{tab_AR(1)_linear}, the BDS test fails to maintain the nominal level of significance, which undermines its reliability. 

For $m=4000$, the RIG-DD test performs equally well or better than the Runs test across the entire considered range of $\rho$. For $m=1000$, the performance of the RIG-DD test is comparable to that of the Runs test, with the RIG-DD test performing better for smaller values of $\rho$.
These findings suggest that the RIG-DD test serves as a more reliable tool for assessing randomness. 

\begin{table}
\caption{Power against AR(1) process for $m$ observations}
    \centering
    \begin{tabular}{ccccccccc}
        \hline
           & $\rho$ & 0.3 & 0.2 & 0.15 & 0.1 & 0.05 & 0.025 & 0 \\
        \hline
        \multirow{4}{5em}{$m=4000$} & \textbf{RIG-EP} & 100 & 99.9 & 96.4 & 71.6 & 25.8 & 9.7 & 5.5 \\
        & \textbf{RIG-DD} & 100 & 100 & 99.8 & 98.4 & 56.0 & 17.3 & 4.4 \\
        & \textbf{Runs} & 100 & 100 & 99.6 & 97.2 & 52.7 & 15.5 & 5.5 \\
        & \textbf{BDS} & 100 &  87.3 & 45.8 & 17 & 8.3 & 6 & 6.4\\
        \hline
        \multirow{4}{5em}{$m=1000$} & \textbf{RIG-EP} & 97.4 & 73.5 & 49.6 & 26.2 & 11.7 & 6.9 & 5.2\\
        & \textbf{RIG-DD} & 100 & 97.8 & 84.2 & 50.8 & 17.5 & 8.1 & 4.7 \\
        & \textbf{Runs} & 100 & 98.0 & 85.6 & 49.7 & 16.6 & 7.5 &  4.5\\
        & \textbf{BDS} & 90.8 & 40.4 & 21.3 & 13.3 & 10.3 & 10.4 & 10.2\\
        \hline
    \end{tabular}
\label{tab_AR(1)_linear}
\end{table}

Building on the strong performance of the RIG-DD test with relatively large datasets, we now examine its effectiveness on smaller datasets, 
which are common in real-world applications where data collection is often challenging or expensive. 
For this, we generate 10000 series of observations from the AR(1) process with sample sizes: (i) $m=120$, (ii) $m=80$, and (iii) $m=40$, using different values of $\rho$. 
The findings of this experiment are presented in Table \ref{tab:AR(1)_small_m}. Notice that the performance of the Runs test is significantly worse for these smaller datasets as compared to the larger ones. For instance, with 4000 observations and $\rho=0.2$, the Runs test achieves near-perfect power (approximately 1), which drops to 0.23 when the number of observations is 120. This illustrates the limitation of the Runs test when faced with smaller datasets. In contrast, the RIG-DD test demonstrates a clear advantage. Across all scenarios, the RIG-DD test outperforms the Runs test, with the performance gap becoming more pronounced as the sample size decreases, especially for smaller values of $\rho$.  This trend underscores the reliability of the RIG-DD test in handling smaller datasets. 
Additionally, while the RIG-EP test also experiences a decline in power, especially for lower values of $\rho$, it remains relatively competitive for higher values of $\rho$. However, the RIG-DD test consistently maintains better performance, highlighting its potential as a superior tool for testing randomness in smaller datasets.

\begin{table}[t]
\caption{Power against AR(1) process for smaller values of $m$.}
    \centering
    \begin{tabular}{cccccccccccc}
        \hline
        & $\rho$ & 0.9 & 0.8 & 0.7 & 0.6 & 0.5 & 0.4 & 0.3 & 0.2 & 0.1 & 0 \\
        \hline
        \multirow{3}{4em}{$m=120$} 
        & \textbf{RIG-EP} & 99.8 & 97.6 & 91.1 & 77.0 & 59.4& 38.8 & 23.7 & 12.6 & 6.0 & 3.2  \\
        & \textbf{RIG-DD} &100 &  100 & 100 & 99.8 & 97.4 & 87.1 & 64.3 & 36.3 &  15.2 & 5.6  \\
        & \textbf{Runs} & 100 & 100 & 99.9 & 98.6 & 92.3 & 76.0 & 49.3 & 23.5& 7.7 & 3.7 \\
        \hline
        \multirow{3}{4em}{$m=80$} & \textbf{RIG-EP} & 98.2 & 93.5 & 83.8 & 70.2 & 53.0 & 37.7 & 24.8 &  14.8 & 8.7 & 5.7 \\
        & \textbf{RIG-DD} & 100 & 99.9 & 99.6 & 96.8 & 88.2 & 70.3 & 46.1 & 25.1 & 11.3 & 5.0 \\
        & \textbf{Runs} & 100 & 99.5 & 97.8 & 91.1 & 76.7 & 55.6 & 32.4 & 15.2 & 6.1 & 3.1 \\
        \hline
        \multirow{3}{4em}{$m=40$} & \textbf{RIG-EP} & 72.6 & 58.5 & 44.5 & 33.4 & 23.4 & 15.4 & 10.0 & 6.4 & 4.6 & 3.7\\
        & \textbf{RIG-DD} & 98.2 & 95.2 & 88.1 & 76.3 & 58.4 & 41.8 & 27.3 & 15.3 & 8.6 & 4.6 \\
        & \textbf{Runs} & 94.7 & 89.1 & 77.0 & 63.3 & 46.0 & 29.9 & 17.6 & 9.1 & 5.0 & 3.3 \\
        \hline
    \end{tabular}
\label{tab:AR(1)_small_m}
\end{table}

In real-world time series data, dependencies often extend beyond a single lag, making ARMA processes well-suited for capturing such patterns. For the next experiment, we use ARMA(2,1) ($Y_t = \rho_1 Y_{t-1} + \rho_2 Y_{t-2} + e_t + \psi e_{t-1}$) process as the underlying data-generating mechanism. 
We generate 1000 series with sample sizes $m=1000$ and $m=160$, setting $\varphi = 0.15$ and varying the parameter vector $\Vec{\rho} = (\rho_1,\rho_2$). All four tests are applied to the resulting datasets, and the outcomes are presented in Table \ref{tab_ARMA}. Additionally, we perform similar experiments with the data-generating mechanism as AR(2) ($Y_t = \rho_1 Y_{t-1} + \rho_2 Y_{t-2} + e_t$) and MA(1) ($Y_t = e_t + \psi e_{t-1}$), results of which are provided in Tables \ref{tab_AR2} and \ref{tab_MA1}, respectively. 

\begin{table}[t]
\caption{Power against ARMA(2,1) process with $\varphi = 0.15$ for $m$ observations}
    \centering
    \begin{tabular}{ccccccc}
        \hline
         & $\Vec{\rho}$ & $(0.2,0.2)$ & $(0.1,-0.1)$ & $(0.05,0.05)$ & $(0.025,-0.025)$ \\
        \hline
        \multirow{4}{4.5em}{$m=1000$} & \textbf{RIG-EP} & 100 & 93.0 & 77.0 & 50.5 \\
        & \textbf{RIG-DD} & 100 & 100 & 98.8 & 84.5 \\
        & \textbf{Runs} & 100 & 100 & 98.4 &  85.4 \\
        & \textbf{BDS} & 100 & 61.4 & 43.8 & 30.4 \\
        \hline
         \multirow{4}{4.5em}{$m=160$} & \textbf{RIG-EP} & 51.0 & 16.7 & 16.9 & 12.0 \\
        & \textbf{RIG-DD} & 92.0 & 46.7 & 43.7 & 34.1\\
        & \textbf{Runs} & 87.4 & 41.0 & 37.0 & 25.4 \\
        & \textbf{BDS} & 65.2 & 38.9 & 36.0 & 36.6\\
        \hline
    \end{tabular}
\label{tab_ARMA}
\end{table}

\begin{table}[h!]
\caption{Power against AR(2) process for $m$ observations}
    \centering
    \begin{tabular}{ccccccc}
        \hline
         & $\Vec{\rho}$ & $(0.2,0.2)$ & $(0.1,-0.1)$ & $(0.05,0.05)$ & (0.025,-0.025) & $(0,0)$ \\
         \hline
        \multirow{4}{5em}{$m=1000$} & \textbf{RIG-EP} & 87.8 & 24.2 & 12.3 & 8.2 & 4.7\\
        & \textbf{RID-DD} & 100 & 47.3 & 17.3 & 9.7 & 5.2\\
        & \textbf{Runs} & 99.2 & 42.8 & 17.1 & 7.8 & 5.3\\
        & \textbf{BDS} & 89.2 & 14.2 & 11.5 & 10.1 & 9.3\\
        \hline
        \multirow{4}{5em}{$m=160$} & \textbf{RIG-EP} & 23.5 & 7.4 & 3.6 & 3.8 & 4.5\\
        & \textbf{RID-DD} & 59.8 & 16.4 & 9.3 & 7.0 & 4.8\\
        & \textbf{Runs} & 52.4 & 8.5 & 7.9 & 4.8 & 4.3\\
        & \textbf{BDS} & 47.5 & 34.1 & 35.7 & 35.4 & 35.4\\
        \hline
    \end{tabular}
\label{tab_AR2}
\end{table}

\begin{table}[h]
\caption{Power against MA(1) process for $m$ observations}
    \centering
    \begin{tabular}{cccccccc}
        \hline
         & $\varphi$ &  0.5 & 0.4 & 0.3 & 0.2 & 0.1 & 0\\
         \hline
        \multirow{4}{5em}{$m=1000$} & \textbf{RIG-EP} & 99.8 & 98.9 & 93.6 & 69.5 & 24.4 & 5.4\\
        & \textbf{RID-DD} & 100 & 100 & 100 & 96.2 & 48.9 & 4.6\\
        & \textbf{Runs} & 100 & 100 & 100 & 97.3 & 46.9 & 5.2\\
        & \textbf{BDS} & 100 & 99.3 & 84.7 & 35.8 & 11.9 & 9.8\\
        \hline
        \multirow{4}{5em}{$m=160$} & \textbf{RIG-EP} & 53.7 & 40.5 & 25.1 & 12.7 & 6.0 & 4.2\\
        & \textbf{RID-DD} &  93.9 & 85.1 & 69.2 & 40.4 & 14.9 & 5.4\\
        & \textbf{Runs} &  91.6 & 81.7 & 60.3 & 30.5 & 10.9 & 5.4\\
        & \textbf{BDS} & 71.6 & 55.8 & 41.3 & 35.0 & 32.9 & 33.7\\
        \hline
    \end{tabular}
\label{tab_MA1}
\end{table}

Tables \ref{tab_ARMA}, \ref{tab_AR2}, and \ref{tab_MA1} illustrate that the RIG-EP and RIG-DD tests consistently outperform the BDS test. Notably, the RIG-DD test demonstrates superior performance for smaller $m$ and performs comparably to the Runs test for larger $m$ across the entire range of $\Vec{\rho}(\psi)$ considered. Across all three processes, AR(2), MA(1), and ARMA(2,1), the RIG-EP test performs competitively with the Runs test at higher values of $\rho$ ($||\Vec{\rho}||$).  Overall, the results highlight the superior and consistent performance of the RIG-DD test compared to both Runs and BDS tests in detecting dependence.

Scenarios where the serial correlation is zero yet the time series remains dependent often arise due to variations in volatility, a phenomenon typically driven by conditional heteroscedasticity. Such dependencies are commonly captured using models like ARCH and GARCH, which focus on characterizing changes in variance. In contrast, ARMA processes model $Y_t$  as a linear function of its past values and past noise terms, assuming homoscedasticity. To evaluate the ability of our tests to identify these kinds of dependence, we now apply them to datasets exhibiting conditional heteroscedasticity.

We generate 1000 series of size $m$ from the ARCH(1) process defined as $e_t = \sigma_t \epsilon_t, \: \sigma_t^2 = 10^{-6}+ \alpha e_{t-1}^2, \epsilon_t \sim N(0,1) $,
with (i) $m=160$, (ii) $m=1000$, and (iii) $m=7000$ and varying values of $\alpha$. We then apply all four tests to these datasets. The outcomes of these simulations are presented in Table \ref{tab:ARCH}.
The Runs test fails to detect dependence, yielding rejection rates close to the nominal level of significance for both small and large datasets. This may be because the correlation is zero for the ARCH(1) process. In contrast, the RIG-DD test performs much better with its power (for $\alpha=0.9$) rising from 0.319 for $m=160$ to 0.959 for $m=1000$, and reaching 1 for $m=7000$. 

Although the BDS test is designed to detect non-linear dependence, its empirical size is excessively high, reaching 31.5\% for $m=160$, and only approaching the nominal 5\% level for $m\geq 7000$. This inflation raises concerns about the reliability of the results obtained from the BDS test. Moreover, for $m \geq 7000$, where the empirical size of the BDS test aligns closer to 5\%, the RIG-DD test demonstrates competitive performance across nearly all values of $\alpha$.
Furthermore, unlike the BDS test, the RIG-DD test consistently maintains the 5\% level of significance under all conditions. These findings establish the RIG-DD test as a more reliable tool for assessing randomness, effectively detecting complex dependencies, including those found in conditional heteroscedastic settings.

\begin{table}[t]
\caption{Power against ARCH(1) process with $m$ observations.}
    \centering
    \begin{tabular}{cccccccccccc}
        \hline
        & $\alpha$ & 0.9 & 0.8 & 0.7 & 0.6 & 0.5 & 0.4 & 0.3 & 0.2 & 0.1 & 0 \\
        \hline
        \multirow{4}{4.5em}{$m=160$} 
        & \textbf{RIG-EP} & 8.8 & 8.1 & 7.1 & 5.6 & 6.2 & 5.8 & 5.0 & 3.5 & 4.6 & 3.8 \\
        & \textbf{RIG-DD} & 31.9 & 28 & 23.8 & 19 & 18.6 & 15.2 & 10.5 & 9.5 & 5.4 & 4.2 \\
        & \textbf{Runs} & 4.9 & 4.6 & 4.8 & 4.0 & 4.7 & 4.1 & 4.2 & 4.7 & 6.1 & 5.9 \\
        & \textbf{BDS}   & 98.7 & 96.7 & 93 & 82.4 & 71.8 & 58.7 & 35.5 & 33.3 & 39.8 & 31.5\\
        \hline
        \multirow{4}{4.5em}{$m=1000$} & \textbf{RIG-EP} & 31.6 & 25.2 & 24.3 & 18.9 & 15.2 & 10.5 & 9.4 & 8.5 & 4.9 & 3.4 \\
        & \textbf{RIG-DD} & 95.9 & 93.6 & 87.4 & 80.1 & 61.7 & 44.0 & 27.8 & 18.1 & 7.5 & 4.4 \\
        & \textbf{Runs} & 3.8 & 5.8 & 4.7 & 4.7 & 4.5 & 3.7 & 6.5 & 4.6 & 5.2 & 5.2 \\
        & \textbf{BDS} & 100 & 100 & 100 & 100 & 100 & 100 & 99.7 & 91.8 & 50.1 & 10.5\\
        \hline
         \multirow{4}{4.5em}{$m=7000$} & \textbf{RIG-EP} & 94.0 & 89.0 & 81.6 & 67.6 & 63.1 & 46.1 & 31.7 & 17.9 & 7.1 & 4.0\\
        & \textbf{RIG-DD} & 100 & 100 & 100 & 100 & 100 & 100 & 99.9 & 91.2  & 33.1 & 5.3\\
        & \textbf{Runs} & 5.8 & 4.4 & 3.9 & 5.8 & 6.6 & 4.4 & 5.0 & 4.2 & 5.0 & 5.8\\
        & \textbf{BDS}  & 100 & 100 & 100 & 100 & 100 & 100 &  100 & 100 & 100 & 5.4\\
        \hline
    \end{tabular}
\label{tab:ARCH}
\end{table}

While ARCH models capture volatility using only past squared errors, the GARCH models include both past squared errors and past variances. We next generate 1000 datasets of size $m$ from the GARCH(1,1) process defined as $e_t = \sigma_t \epsilon_t, \: \sigma_t^2 = 10^{-6} + \alpha e_{t-1}^2 + \beta \sigma_{t-1}^2, \epsilon_t \sim N(0,1)$ with (i) $m=160$, (ii) $m=1000$ and (iii) $m=7000$, and varying $(\alpha,\beta)$. We then apply all four tests on these datasets. The outcomes of these simulations are presented in Table \ref{tab:GARCH}.
Like the ARCH process, the Runs test again fails to detect dependence in the GARCH process. However, the RIG-DD test demonstrates stronger power of $0.334$ for $m=160$, increasing to 0.948 for $m=1000$, and reaching 1 for $m=7000$.
Similar to the ARCH case, the BDS test struggles with inflated empirical sizes, reaching close to 5\% only for $m \geq 7000$. Furthermore, when the empirical size of the BDS test is approximately 5\% level at $m = 7000$, the RIG-DD test performs competitively for almost all considered values of ($\alpha,\beta$). Given its ability to maintain the level of significance while effectively capturing dependence, the RIG-DD test proves to be a more reliable tool for detecting dependencies of the GARCH type.

\begin{table}[t]
\caption{Power against GARCH(1,1) process with $m$ observations.}
    \centering
    \begin{tabular}{ccccccccc}
        \hline
       & $(\alpha,\beta)$ & (0.8,0.1) & (0.7,0.2) & (0.5,0.2) & (0.2,0.7)  & (0.2,0.5) & (0.1,0.8) & (0,0) \\
        \hline
        \multirow{4}{4.5em}{$m=160$} & \textbf{RIG-EP} & 8.1 & 7.1 & 5.2  & 5.5 & 4.1  & 4.7 & 4.4 \\
        & \textbf{RIG-DD} & 33.4 & 29.5 & 20.5 & 13.5 & 11.1  & 8.8 & 4.7 \\
        & \textbf{Runs}   & 4.3 & 4.5 & 5.0 & 4.3 & 3.3 & 5.2 & 5.5 \\
        & \textbf{BDS}  & 98.6 & 97.6 & 90.1 & 74.3 & 64.8  & 49.1 & 32.2 \\
        \hline
        \multirow{4}{4.5em}{$m=1000$} & \textbf{RIG-EP} & 32.1 & 27.9 & 16.9 & 10.1 & 7.4  & 6.4 & 5.1 \\
        & \textbf{RIG-DD} & 94.8 & 93.8 & 71.8 & 28.7 & 21.7  & 10.9 & 5.2 \\
        & \textbf{Runs}   & 4.2 & 4.8 & 4.0 & 5.3 & 3.7  & 5.2 & 4.2\\
        & \textbf{BDS}  & 100 & 100 & 100 & 100 & 99.5  & 94.9 & 9.7\\
        \hline
        \multirow{4}{4.5em}{$m=7000$} & \textbf{RIG-EP} & 95.8 & 94.4 & 71.7 & 33.4 & 21.7 & 12.6 & 4.0\\
        & \textbf{RIG-DD} & 100 & 100 & 100 & 99.8 & 96.2 & 58.6 & 5.3 \\
        & \textbf{Runs}   & 5.8 & 5.3 & 4.3 & 5.3 & 5.6 & 5.2 & 5.8 \\
        & \textbf{BDS}  & 100 & 100 & 100 & 100 & 100 &100 & 5.4\\
        \hline
    \end{tabular}
\label{tab:GARCH}
\end{table}

Next, we turn our attention to more complex processes that may exhibit chaotic behavior. While chaos and randomness can both produce irregular and seemingly unpredictable patterns, they differ fundamentally in nature. Chaos originates from deterministic systems governed by well-defined mathematical rules, where future states are fully determined by initial conditions. However, due to their extreme sensitivity to these initial conditions, even small differences can lead to drastically different outcomes over time, making long-term prediction practically impossible. As a result, chaotic systems may appear random, but they are inherently deterministic \citep{chaos}. Given this distinction, we investigate whether our proposed tests can differentiate chaotic behavior from true randomness. For this, we conduct simulations using a well-known chaotic system, the Logistic Map. We apply the four previously used dependence tests along with the 0-1 test for Chaos, which is specifically designed to detect chaos in data.

We generate 1000 series from the Logistic Map process, defined as $x_{n+1} = a x_n (1-x_n) + e_t, e_t \sim N(0,s)$ for various values of $a$ by fixing $x_0=0.4$ and $s=0.001$. To minimize the influence of the initial condition and allow the process to settle into its long-run behavior, we discard the first 1000 observations. We then take the subsequent $m$ observations for testing. The results are presented in Table \ref{tab_chaos}.
When $a=4$, the system exhibits fully developed chaos. In this case, the Runs test fails completely, yielding nearly 0 power, while the RIG-DD test successfully detects the chaos with very high power (nearly 1). Additionally, the BDS test and the 0-1 test for Chaos also exhibit high power, whereas the RIG-EP test does not perform well.
For values of $a$ in the range $3.57<a<4$, the process remains dependent, and chaos begins to emerge but is not fully developed. In this range, all tests detect the dependence with power nearly 1. For $3<a<3.57$, the process is strongly non-linear and dependent, but not chaotic. In this case, the 0-1 test for Chaos shows nearly 0 power, correctly indicating the absence of chaos, while the other tests detect the underlying dependence with approximately 1 power. For $0<a<3$, the system remains dependent but converges to a stable fixed point. Here, the Runs, BDS, and RIG-EP tests fail to detect the dependence, while RIG-DD detects it with power nearly 1. These findings demonstrate the reliability and effectiveness of the RIG-DD tests in detecting a broad spectrum of dependence structures, including non-linear, chaotic, and non-chaotic forms, with very high power across all scenarios.

\begin{table}[t]
\caption{Power against Logistic Map process with $m$ observations.}
    \centering
    \begin{tabular}{ccccccccc}
        \hline
        & $a$ & 2.9 & 3.5 & 3.8 & 4\\
        \hline
        \multirow{5}{4.5em}{$m=160$} & \textbf{RIG-EP} & 4.6 & 100 & 100 & 1.3\\
        & \textbf{RIG-DD} &100 & 100 & 100 & 100 \\
        & \textbf{Runs} & 4.8 & 100 & 100 & 0\\
        & \textbf{BDS} & 21.2 & 100 & 100 & 100 \\
        &  \textbf{Chaos 0-1} & 0 & 0 & 100 & 100 \\
        \hline
        \multirow{5}{4.5em}{$m=1000$} & \textbf{RIG-EP} & 4.8 & 100 & 100 & 4.6\\
        & \textbf{RIG-DD} &100 & 100 & 100 & 100 \\
        & \textbf{Runs} & 4.6 & 100 & 100 & 0\\
        & \textbf{BDS} & 6.0 & 100 & 100 & 100 \\
        &  \textbf{Chaos 0-1} & 0 & 0 & 100 & 100 \\
        \hline
     \end{tabular}
\label{tab_chaos}
\end{table}

Until now, our analysis focused on datasets where the number of observations, $m$, was a multiple of four for the RIG-EP test and a multiple of two for the RIG-DD test. To further assess the reliability and effectiveness of our approach, we extended our evaluation to cases where $m$ does not follow these constraints. This will provide us with insights into the performance and reliability of the RIG-EP test in more diverse and practical data conditions.

First, for the RIG-EP test, we generate $m$ observations from an AR(1) process with varying values of $\rho$. We considered two cases:(A) for larger datasets having values of $m$ as (i) $m = 2000$, (ii) $m = 2001$, (iii) $m = 2002$, and (iv) $m = 2003$ and (B) for smaller dataset having the values of $m$ as (v) $m=160$, (vi) $m=161$, (vii) $m=162$ and (viii) $m=163$. We apply the RIG-EP test to each of these datasets, assessing randomness at the 5\% level of significance. For cases where the number of observations was not a multiple of four, we implement Algorithm \ref{algo_RIG-EP} using Benjamini-Hochberg corrections.  The results of this experiment are presented in Tables \ref{tab:not_multiple_of_4_1} and \ref{tab:not_multiple_of_4_2}. The outcomes for cases (ii), (iii), and (iv) are closely aligned with those of case (i), demonstrating that the method remains effective even when the number of observations is not a multiple of four. A similar thing happens for smaller datasets as well.

\begin{table}[t]
\caption{Power comparison of RIG-EP Test against AR(1) process for larger datasets when $m$ is and is not a multiple of four.}
    \centering
    \begin{tabular}{ccccccccc}
        \hline
        $\rho$ & 0.3 & 0.2 & 0.15 & 0.1 & 0.05 & 0.025 & 0 \\
        \hline
        $m=2000$ &  100 & 93.6 & 74 & 39.7 & 14.4 & 7.5 & 4.7 \\
        $m=2001$ &  100 & 99.3 & 88.4 & 55.2 & 19.9 & 7.4 & 4.1 \\
        $m=2002$ & 100 & 99.6 & 93.8 & 62.3 & 21.2 & 7.6 & 5.4 \\
        $m=2003$ & 100 & 99.8 & 94.5 & 63.3 & 18.4 & 8.5 & 5.6 \\
        \hline
    \end{tabular}
\label{tab:not_multiple_of_4_1}
\end{table}

\begin{table}[t]
\caption{Power comparison of RIG-EP Test against AR(1) process for smaller datasets when $m$ is and is not a multiple of four.}
    \centering
    \begin{tabular}{ccccccccc}
        \hline
        $\rho$ & 0.7 & 0.6 & 0.5 & 0.4 & 0.3 & 0.2 & 0.1 & 0 \\
        \hline
        $m=160$ &  96.8 & 90.4 & 75.9 & 52.2 & 31.1 & 16.5 & 6.8 & 3.9 \\
        $m=161$ &  99.9 & 96 & 85.3 & 64.5 & 37.1 & 19.1 & 6.9 & 3.7 \\
        $m=162$ & 99.9 & 99.1 & 93.7 & 75.7 & 48.6 & 26.6 & 9.9 & 5.0 \\
       $m=163$ &  100 & 99.7 & 94.2 & 82.2 & 55.8 & 28.5 & 12.4 & 4.4 \\
        \hline
    \end{tabular}
\label{tab:not_multiple_of_4_2}
\end{table}

For the RIG-DD test, we generate $m$ observations from an AR(1) process with different values of $\rho$. We consider two values of $m$: $m = 1000$ and $m = 1001$. We apply the RIG-DD test to each of these datasets as given in Algorithm \ref{algo_RIG_DD}, assessing randomness at the 5\% level of significance. The results of this experiment are presented in Table \ref{tab:not_multiple_of_2}. The simulation findings indicate that the findings for $m=1000$ and $m=1001$ are closely aligned, demonstrating that the method remains effective even when the number of observations is not a multiple of two. 

\begin{table}[t]
\caption{Power comparison of RIG-DD Test against AR(1) process when $m$ is and is not a multiple of two.}
    \centering
    \begin{tabular}{ccccccccccc}
        \hline
        $\rho$ & 0.3 & 0.2 & 0.15 & 0.1 & 0.05 & 0 \\
        \hline
        $m=1000$ & 100 & 98.8 & 84 & 52.2 & 17.9 & 5.7\\
        \hline
        $m=1001$  & 100 & 99.7 & 94.2 & 58.7 & 17.5 & 3.7 \\
        \hline
    \end{tabular}
\label{tab:not_multiple_of_2}
\end{table}

When the data is actually random, any permutation of that data will also be random.
Since the construction of the RIG depends on the specific sequence of the observations, it is important to check whether, when the data is actually random, any permutation of the sequence affects the level of the tests. Therefore, we assess the performance of our tests against permuted sequences of random series. 
For this, we generate a set of 1000 random observations from (i) the Uniform distribution, (ii) the Normal(0,1) distribution, and (iii) the Cauchy(0,1) distribution, and create 1000 permutations for each. We test for randomness of each of these permuted series at 5\% level of significance. The RIG-EP test rejected the null hypothesis for 5.7\%, 5.3\%, and 4.8\%  permutations from (i), (ii), and (iii), respectively. The RIG-DD test showed rejection rates of 6.1\%, 4.3\%, and 5.7\% permutations from (i),(ii), and (iii), respectively. The results show that the level is approximately maintained in all cases, confirming the reliability of our tests. 

\section{Real World Applications} 
\label{real_data}


In this section, we apply the proposed RIG-EP and RIG-DD tests to two real-world datasets: the quarterly growth rate of the U.S. Gross National Product and the daily stock returns of IBM. The analysis of these datasets is detailed as follows.

\subsection{Quarterly growth rate of US GNP} 

Consider the quarterly growth rate of the U.S. real gross national product (GNP), seasonally adjusted, from the second quarter of 1947 to the first quarter of 1991 \citep{time_series}. The data has 176 entries.
Applying the RIG-EP test, we find  $|\hat{p}-\frac{1}{3}| =  0.2576$, which exceeds the cutoff $c=0.1393$ at the 5\% level of significance, leading us to reject the null hypothesis of randomness. 
Now applying the RIG-DD test, we note that the value of the test statistic is 1.3604, which surpasses the threshold of 0.7192 for the 5\% level of significance. Consequently, we reject the null hypothesis of randomness. Thus, both the RIG-EP and RIG-DD tests indicate the presence of dependence in the GNP dataset. The Runs test also indicates the presence of dependence with a p-value of 0.00815.

\subsection{Daily simple returns of stock} 

In most applications in finance, the daily returns are conventionally assumed to be mutually independent \citep{time_series}. The implications of this hypothesis extend far and wide, influencing investment strategies, portfolio management, and market behaviour. Therefore, checking for the randomness of the data is essential to avoid financial pitfalls. 

For our analysis, we utilize the daily simple returns of IBM, which comprises 9845 entries collected from January 2, 1970, to December 31, 2008 \citep{time_series}. After removing two missing values, we conduct the analysis on the remaining 9843 entries. It is observed that the data has high kurtosis. 
Applying the RIG-DD test (Algorithm \ref{algo_RIG_DD}), we obtain the test statistics as 0.8591 and 1.1452, both of which exceed the threshold value of 0.6516, leading to the rejection of the null hypothesis of randomness at the 5\% level of significance.
In contrast, the RIG-EP test fails to reject the null hypothesis at the same level of significance. 

As discussed in Section \ref{intro}, the ACF plots of squared and absolute returns suggest the possibility of dependence in the data. To validate the findings of the RIG-DD test, we apply the Lagrange Multiplier test for conditional heteroskedasticity in the data, which provides strong evidence for it, with a highly significant p-value of $2.2 \times 10^{-16}$. The BDS test also supports this conclusion, yielding a p-value below 0.0001. On the other hand, the Runs test fails to detect any dependence, producing a non-significant p-value of 0.3973.

These results demonstrate that the RIG-DD test effectively captures complex dependencies such as conditional heteroskedasticity and nonlinearity that may not be detected through usual autocorrelation-based methods. By identifying such hidden structures, the RIG-DD test provides a more comprehensive and sensitive tool for assessing randomness in data.


\section{Conclusion}
\label{conclusions}

Statistical and financial models often rely on the critical assumption that data is generated randomly. However, there are not many parametric and non-parametric techniques available to validate the randomness assumption in the data. In this paper, we use the properties of RIG, which are independent of the underlying distribution of observations and rely solely on the randomness of data. This allows our proposed tests to be broadly applicable across various types of data and distributional contexts. 
We introduced two tests: the RIG-Edge Probability (RIG-EP) and RIG-Degree Distribution (RIG-DD).

Our extensive simulation studies confirm the reliability and effectiveness of the proposed tests. The RIG-DD test consistently outperforms both the Runs and BDS tests across a range of scenarios. Furthermore, both RIG-EP and RIG-DD tests successfully detect dependencies beyond simple correlation, including conditional heteroskedasticity and chaotic processes, highlighting their ability to capture more complex patterns of dependence in data. The practical utility of our tests is further showcased through applications to real-world datasets, such as economic indicators and stock returns.

\section{Data Availability Statement}\label{data-availability-statement}

All datasets used in this paper are taken from \cite{time_series} and are openly available at https://faculty.chicagobooth.edu/ruey-s-tsay/research/analysis-of-financial-time-series-3rd-edition. The US GNP data can be found in the file 'dgnp82.txt', and the IBM simple returns data is provided in 'd-ibm3dx7008.txt', on the same website.

\appendix

\section{Threshold Values for RIG-DD Test}

Table \ref{tab_cutoff_linear} presents the threshold values ($C_\alpha$) for the RIG-DD test at the level of significance $\alpha$, based on $m$ observations, as computed using Algorithm \ref{algo_thrsld}.

\begin{table}[h!]
\caption{Threshold values for RIG-DD test at different levels of significance with $m$ observations.}
    \centering
    \begin{tabular}{cccc||cccc}
        \hline
        \multirow{2}{*}{$m$} & \multicolumn{3}{c||}{Level of Significance} & \multirow{2}{*}{$m$} & \multicolumn{3}{c}{Level of Significance} \\
        & 10\%  & 5\%  & 1\%  & & 10\%  & 5\%  & 1\% \\
        \hline
        50 & 0.6638 & 0.7686 & 0.9308 & 1300 & 0.6027 & 0.6718 & 0.8543 \\
        100  & 0.6012 & 0.6818 & 0.9507 & 1400 & 0.5692 & 0.6547 & 0.8248 \\
        120  & 0.6325 & 0.6972 & 0.9272 & 1500 & 0.5775 & 0.6539 & 0.7988 \\
        140  & 0.6264 & 0.7130 & 0.9308 & 1600 & 0.6014 & 0.6967 & 0.8725 \\
        160  & 0.5950 & 0.6979 & 0.8532 & 1700 & 0.6061 & 0.7142 & 0.9066 \\
        180  & 0.6054 & 0.7049 & 0.9630 & 1800 & 0.6118 & 0.7139 & 0.9513 \\
        200  & 0.6159 & 0.7036 & 0.8900 & 1900 & 0.5821 & 0.6612 & 0.8310 \\
        240  & 0.6122 & 0.7141 & 0.9051 & 2000 & 0.5875 & 0.6736 & 0.8429 \\
        280  & 0.6173 & 0.6950 & 0.8504 & 2500 & 0.5617 & 0.6715 & 0.8050 \\
        320  & 0.6155 & 0.7021 & 0.8608 & 2800 & 0.5817 & 0.6724 & 0.8783 \\
        360  & 0.5807 & 0.6885 & 0.8754 & 3000 & 0.5685 & 0.6866 & 0.8209 \\
        400  & 0.6030 & 0.7059 & 0.9198 & 3500 & 0.5815 & 0.6727 & 0.8335 \\
        500  & 0.6182 & 0.7040 & 0.8706 & 3800 & 0.5918 & 0.6849 & 0.8341 \\
        600  & 0.5947 & 0.6844 & 0.8838 & 4000 & 0.5889 & 0.6729 & 0.8384 \\
        700  & 0.5849 & 0.6625 & 0.8427 & 4500 & 0.5888 & 0.6762 & 0.8822 \\
        800  & 0.5801 & 0.6651 & 0.9191 & 5000 & 0.5815 & 0.6750 & 0.8289 \\
        900  & 0.5953 & 0.7005 & 0.8666 & 5500 & 0.5606 & 0.6665 & 0.8421 \\
        1000 & 0.5885 & 0.6989 & 0.8594 & 5800 & 0.6190 & 0.7061 & 0.8478 \\
        1100 & 0.5799 & 0.6731 & 0.8888 & 6000 & 0.5797 & 0.6814 & 0.8397 \\
        1200 & 0.6243 & 0.6954 & 0.9498 & 7000 & 0.5825 & 0.6592 & 0.8543 \\
        \hline
    \end{tabular}
\label{tab_cutoff_linear}
\end{table}


\bibliography{ref}

\end{document}